\documentclass[prb,twocolumn,showpacs,preprintnumbers,amsmath,amssymb,superscriptaddress]{revtex4-1}
\usepackage{graphicx}
\usepackage{dcolumn}
\usepackage{bm}
\usepackage{dsfont}
\usepackage{color}

\begin{document}

\title{Statistical analysis of the figure of merit of a two-level thermoelectric system: a random matrix approach} 
\author{A. Abbout}\email{abbout.adel@gmail.com}
\affiliation{Physical Science and Engineering Division, King Abdullah University of Science and Technology (KAUST), Thuwal 23955-6900, Saudi Arabia}
\author{H. Ouerdane}
\affiliation{Russian Quantum Center, 100 Novaya Street, Skolkovo, Moscow Region 143025, Russia}
\affiliation{UFR Langues Vivantes Etrang\`eres, Universit\'e de Caen Normandie, Esplanade de la Paix 14032 Caen, France}
\author{C. Goupil}
\affiliation{Laboratoire Interdisciplinaire des Energies de Demain, LIED/CNRS UMR 8236 Universit\'e Paris Diderot; B\^at. Lamarck B 35 rue H\'el\`ene Brion 75013 Paris}
\date{\today}

\begin{abstract}
Using the tools of random matrix theory we develop a statistical analysis of the transport properties of thermoelectric low-dimensional systems made of two electron reservoirs set at different temperatures and chemical potentials, and connected through a low-density-of-states two-level quantum dot that acts as a conducting chaotic cavity. Our exact treatment of the chaotic behavior in such devices lies on the scattering matrix formalism and yields analytical expressions for the joint probability distribution functions of the Seebeck coefficient and the transmission profile, as well as the marginal distributions, at arbitrary Fermi energy. The scattering matrices belong to circular ensembles which we sample to numerically compute the transmission function, the Seebeck coefficient, and their relationship. The exact transport coefficients probability distributions are found to be highly non-Gaussian for small numbers of conduction modes, and the analytical and numerical results are in excellent agreement. The system performance is also studied, and we find that the optimum performance is obtained for half-transparent quantum dots; further, this optimum may be enhanced for systems with few conduction modes.
\end{abstract}
\maketitle

\section{Introduction}
Low-dimensional systems offer a wealth of technological possibilities thanks to the rich variety of artificial custom-made semiconductor-based structures that nowadays may be routinely produced. The properties of these systems, including confinement geometry, density of states, and band structure, may be tailored on demand to control the transport of heat and the transport of confined electrical charges, as well as their coupling. In the current context of intense research in energy conversion physics, this is crucial to further improve the performance and increase the range of operation of present-day thermoelectric devices. These latter are characterized by their so-called figure of merit, which, in the frame of linear response theory, may be expressed as: $ZT=\sigma s^2 T/\kappa$, where $s$ is the Seebeck coefficient, $T$ is the average temperature across the device, and $\sigma$ and $\kappa$ are the electrical and thermal conductivities respectively, with $\kappa=\kappa_{\rm e}+\kappa_{\rm lat}$, accounting for both electron and lattice thermal conductivities.

Amongst the various low-dimensional thermoelectric systems that have been studied, quantum dots, which are confined electronic systems whose size and shape are controlled by external charged gates, keep attracting much interest because of their narrow, Dirac-like, electron transport distribution functions \cite{Mahan} or, equivalently, sharply peaked energy-dependent transmission profiles $\mathcal{T}$, which permit obtainment of extremely high values of electronic contribution to $ZT$ \cite{Trocha}. There is a simple relationship between the electrical conductivity $\sigma$ appearing in the definition of the figure of merit and the transmission function $\mathcal{T}$: $\sigma=(2 e^2/h) \mathcal{T}$, the proportionality factor being the quantum of conductance ($e$ being the electron charge, and $h$ being Planck's constant).

Models of quantum dots form two broad categories, namely interacting and noninteracting models. Those accounting for electron-electron interactions permit analysis of a rich variety of physical phenomena governed by electronic correlations like, e.g., Coulomb blockade and the related conductance oscillations vs. gate voltage \cite{Datta2004}, peak spacing distributions \cite{Baranger2001}, and phase lapses of the transmission phase \cite{Karrasch}. The use of non-interacting model systems for quantum dots or very small electronic cavities may be justified if these are strongly coupled to the reservoirs, i.e. if the confinement yields a mean level spacing that is large compared to the charging energy $e^2/C$, $C$ being the capacitance of the dot \cite{Brouwer,Alhassid}. So, though noninteracting dot models may be seen as toy models, these may provide in some cases a number of insightful results without the need to resort to advanced numerical techniques such as, e.g., the density matrix numerical renormalization group \cite{Karrasch,Weichselbaum2007}. This is illustrated by, e.g., the resonant level model of thermal effects in a quantum point contact \cite{Adel1}, Fano resonances in the quantum dot conductance \cite{Goldstein2007}, and simple models of phase lapses of the transmission phase \cite{Oreg2007}.

Interesting physics problems in a quantum dot also stem from the chaotic dynamics that may be triggered because of structural disorder, or as the quantum dot itself behaves as a driven chaotic cavity because its shape varies as one of the gates generates a random potential. In these cases, as explained in Ref.~\cite{Shankar2008}, one may be interested in the statistics of the system's spectrum rather than the detailed description of each level. Random matrix theory \cite{Guhr1998} provides tools of choice for this purpose; and, for quantum dots in particular, one may construct a mathematical ensemble of Hamiltonians that satisfies essentially two constraints: the Hamiltonians belong to the same symmetry class and they have the same average level spacing, while the density of states must be the same for the physical ensemble of quantum dots \cite{Shankar2008}. In the context of thermoelectric transport, measurements of thermopower and analysis of its fluctuations~\cite{Godijn} based on the random matrix theory of transport \cite{Beenakker} demonstrated the non-Gaussian character of the distribution of thermopower fluctuations.

In this article, we concentrate on the statistical analysis of the thermoelectric properties of noninteracting quantum dots connected to two leads that serve as electron reservoirs set at different temperatures and electrochemical potentials. We analyze the statistics of $ZT$ in a two-level system connected to two electronic reservoirs set to two slightly different temperatures so that the temperature difference, $\Delta T$, is small enough to remain in the linear response regime: the voltage induced by thermoelectric effect, $\Delta V$, is given by $s=-\Delta V/\Delta T$. The mutual dependence of the transport coefficients that define $ZT$ (e.g., the Wiedmann-Franz law for metals) raises the problem of finding which configuration of the mesoscopic system may yield the largest values of $ZT$, and hence offer optimum performance.

The two-level model presented in this article is the minimal model pertinent for the description of a cavity with two conducting modes presenting a \emph{completely chaotic} behavior \cite{Abbout2}. We consider Fermi energies lying in the vicinity of the spectrum edge of the system's Hamiltonian. This situation is in contrast with the case of high level cavities for which the typical transmission profiles vary much and exhibit numerous maxima and extinctions, which precludes an analytical formulation of the Seebeck coefficient probability distribution since the Cuttler-Mott formula does not apply. Although one may consider a very low temperature regime to concentrate only on a small interval of energies, the need of a large number of levels to ensure the properties of the bulk universality in chaotic systems makes this window of energies so small \cite{Explanation0} that the corresponding temperature tends to become insignificant. Conversely, at the edge of the Hamiltonian spectrum, the typical profile of the transmission is smooth \cite{Explanation} and the analytical treatment of the probability distribution of thermopower is much easier \cite{Abbout1}. Indeed, at this class of universality (spectrum edge) the description of this kind of systems may rest on the \emph{equivalent} minimal chaotic cavity \cite{Abbout2} defined as the system with a number of levels $N$ equal to the number of the conducting modes $M$. By ``equivalent'' we mean that, at the spectrum edge, the original system and the corresponding minimal chaotic cavity lead to the \textit{same statistics} though they certainly have different results for one realization.

The article is organized as follows: In the next section, we introduce the model, the main definitions and notations we use throughout the paper. In Sec. III, we present two derivations of the Seebeck coefficient in the scattering matrix framework: one is general, while the other is made under the restrictive assumption of left-right spatial symmetry. The formal identity of the two results serves as a basis for the statistical analysis presented in Sec. IV, which is the core of the article. We analyze the numerical statistical results obtained for the probability distributions of the figure of merit, thermopower, and power factor. In Sec. V, we analyze the relationship of the Seebeck coefficient to the density of states of the system. In Sec. VI, we extend the discussion to the case of a lattice. The article ends with a discussion and concluding remarks. A Supplemental Material detailing some of the numerical aspects of the work as well as some derivations, accompanies the article \cite{SupMat}. 

\section{Model}
\begin{figure}
\includegraphics[scale=0.275]{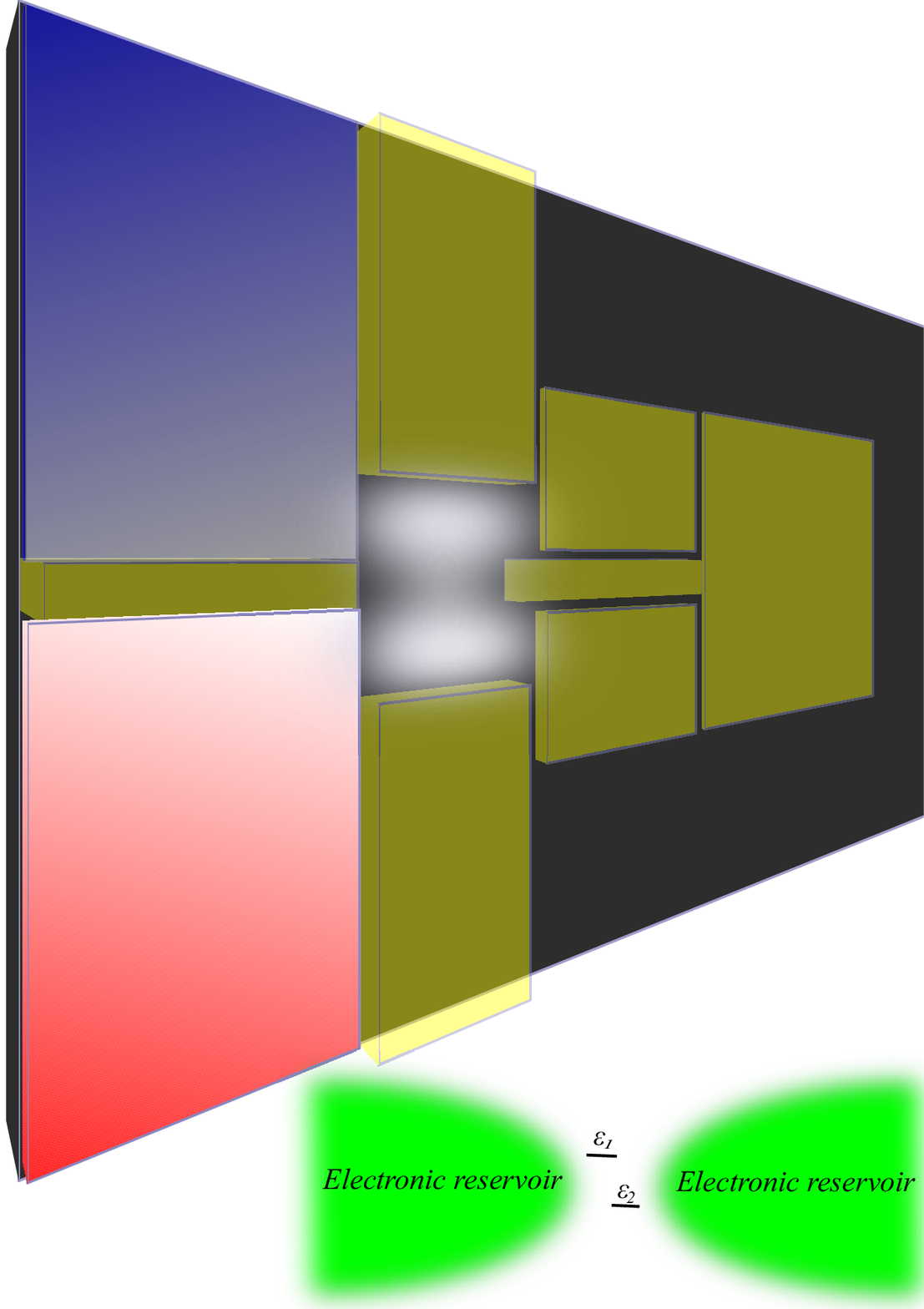}
\caption{(Color online) Schematic figure of the two level-system consisting of negatively charged top gates (yellow) on a two-dimensional electron gas. The red (blue) part represents the hot (cold) electronic reservoir connected to the cavity. The inset shows the system as modeled by the Hamiltonian.}
\label{QDchcav}
\end{figure}

We consider a low-density-of-states two-level quantum dot depicted in Fig.~\ref{QDchcav}. The potential on the right three top gates may be varied randomly in order to slightly modify the shape of the cavity and therefore obtain a statistical ensemble \cite{Godijn}.

\subsection{Preliminary remarks on the minimal chaotic cavities}
Studies of confined systems may be achieved without loss of pertinence by means of simple models, which capture the essential physics of actual systems. A chaotic cavity with $N$ energy levels is modeled with an $N\times N$ Hamiltonian. In the frame of random matrix theory, to obtain a scattering matrix from circular ensembles, the Hamiltonian is distributed from Lorentzian ensembles; other distributions, e.g., the Gaussian distribution, may lead to the same results only for large $N$ in the bulk spectrum. For a fixed value of $N$, if the Fermi energy approaches the bottom of the conduction band (continuum limit) we enter a new universality class: \emph{the edge of the Hamiltonian spectrum} \cite{Abbout1}. The key point here, which will prove extremely useful as shown in the present article, is that at the edge of the spectrum, \emph{all} the distributions obtained with an $N\times N$ Hamiltonian may also be found by considering a simpler case with $M\times M$ matrices, with $M$ ($<N$) being the number of conducting modes. This type of cavity is called the \emph{minimal chaotic cavity}, and permits to lower $N$ down to $N=M$ to simplify the mathematical and computational treatment of the problem, and derive physically meaningful results. As we show, it allows obtainment of very simple formulas for the transport coefficients because the Hamiltonian and the scattering matrix then have the \emph{same} size. It is important because, at the edge of the spectrum (large level spacing), it gives the same result as for the original system ($N>M$). This approach is justified for probing the spectrum edges of mesoscopic systems \cite{Abbout1,Abbout2,AdelPeng} where the physics is different from that of the bulk spectrum \cite{Abbout1,VanLangen}.

\subsection{Scattering matrix}
The two-level system tight-binding Hamiltonian reads as the sum of three contributions: $\mathcal{H}=\mathcal{H}_{\ell}+\mathcal{H}_{\rm s}+\mathcal{H}_{\rm c}$, where $\mathcal{H}_{\ell}=\sum_k \epsilon_k c_k^\dagger c^{\phantom{\dagger}}_k$ is the Hamiltonian of the two (i.e., left and right) leads; $c_k^\dagger$ and $c_k$ are the second-quantized electron creation and annihilation operators in the state $k$. The two energy levels and their coupling are characterized by the Hamiltonian $\mathcal{H}_{\rm s}$:
\begin{equation}
 \mathcal{H}_{\rm s}= \left(
  \begin{array}{ c c }
     v_1 & v_{12} \\
     v_{21} & v_2
  \end{array} \right)
\end{equation}
where the random on-site potentials $v_1$, $v_2$ and coupling $v_{12}$ describe the chaotic behavior of the system. We assume that the left (right) lead is only connected to the site $v_1$ ($v_2$). The contribution ${\mathcal H}_{\rm c}$ involves the coupling matrix $\mathcal{W}$ entering the definition of the scattering matrix \cite{Weidenmuller}:
\begin{equation}
 {\mathcal S}=1-2\pi i \mathcal{W}^\dagger \frac{1}{\epsilon-\mathcal{H}_{\rm s}-\Sigma} \mathcal{W} \label{SandHs}
\end{equation}

\noindent where $\Sigma$ is the self-energy of the two leads. To keep our analysis on a general level, we only give its form: $\Sigma=\Lambda-i\pi  \mathcal{W} \mathcal{W}^\dagger $, with $\Lambda$ being the real part. With the assumption of symmetric reservoirs, the self-energy is proportional to the identity, in which case $ \mathcal{W}=\sqrt{\Gamma/2\pi} {\mathds 1}_{2\times2}$, where $\Gamma=-2\Im\Sigma$ characterizes the broadening. Throughout the article, we adopt the same notations for scalars and their corresponding matrix form when this latter is proportional to the identity.

\section{Derivation of the Seebeck coefficient}

It is instructive to derive an analytic expression of the Seebeck coefficient under the restrictive assumption of spatial left-right symmetry, on the one hand, and compare the result to the that obtained assuming arbitrary energy and arbitrary leads.

\subsection{Left-right spatial symmetry}

The scattering matrix ${\mathcal S}$ is unitary. In the absence of a magnetic field, time reversal symmetry is preserved and the matrix ${\mathcal S}$ is therefore symmetric, which implies $v_{12}=v_{21}$. Moreover, for simplicity, we consider in the first part of this article, systems with left-right spatial symmetry ($v_1=v_2$), so that the reflection from left is the same as that from right. The matrix ${\mathcal S}$ may thus read:
\begin{equation}
  {\mathcal S}=\left(
  \begin{array}{ c c }
     r & t \\
     t & r
  \end{array} \right)
\end{equation}
\noindent where $r$ and $t$ are the reflection and transmission amplitudes respectively. With these definitions, the transmission of the system and the Seebeck coefficient read:
\begin{equation}
 \mathcal{T}=|t|^2, ~~\mbox{and}~~ s=\frac{\partial \ln(\mathcal{T})}{\partial \epsilon} \label{Seebeck}
\end{equation}
The Seebeck coefficient is obtained at low temperatures with the Cutler-Mott formula \cite{CutlerMott}. Here, it is expressed in units of $\frac{\pi^2}{3 e} k^2_{\rm B} T$ (ommiting the sign). The transmission of the system can then directly be obtained using the Fisher-Lee formula \cite{FisherLee}:
\begin{equation}
\mathcal{T}=\Gamma G_{12}\Gamma G_{12}^\dagger
\end{equation}
where the off-diagonal element of the Green's matrix is $G_{12}=\frac{1}{2}\frac{(\epsilon_1-\epsilon_2)}{(\epsilon-\epsilon_1-\Sigma)(\epsilon-\epsilon_2-\Sigma)} $, with $\epsilon_1$ and $\epsilon_2$ being the eigenvalues of $\mathcal{H}_{\rm s}$. We may thus write: 
\begin{equation}
\mathcal{T}(\epsilon)=\frac{\Gamma^2}{4} \frac{(\epsilon_1-\epsilon_2)^2}{|(\epsilon-\epsilon_1-\Sigma)(\epsilon-\epsilon_2-\Sigma)|^2 } \label{Transmission0}
\end{equation}
\noindent which is valid if the left/right symmetry condition is satisfied. Now, combination of Eqs. (\ref{Seebeck}) and (\ref{Transmission0}) yields an expression of the Seebeck coefficient containing the scattering matrix ${\mathcal S}$:
\begin{equation}\label{Seebeck2}
 s=\frac{\alpha}{2}{\rm Tr}\left(\frac{e^{i\Theta}{\mathcal S}- e^{-i\Theta}{\mathcal S}^\dagger}{2i}\right)
\end{equation}
\noindent with $\alpha=4|1-\dot{\Sigma}|/\Gamma$, $\Theta={\rm Arg}(1-\dot{\Sigma})$, and $\dot{\Sigma}\equiv\partial_{\epsilon}\Sigma$. For an energy $\epsilon=\epsilon_0$, which corresponds in general to the middle of the conduction band of a semi-infinite lead (or half-filling limit), the imaginary part of the self-energy derivative vanishes:  $\Im\dot{\Sigma}(\epsilon_0)=0$.

\subsection{General derivation}

The starting point is the transmission of the two-level system:
\begin{equation}
 \mathcal{T}=\Gamma^2\frac{v^2_{12}}{|(\epsilon-\epsilon_1-\Sigma)(\epsilon-\epsilon_2-\Sigma)|^2}
\end{equation}
This formula applies both in presence or absence of the left/right spatial symmetry. The Seebeck coefficient in units of $\frac{\pi^2}{3} \frac{(k_{\rm B}^2 T)}{e}$ reads: $s=\partial_\epsilon \ln(\mathcal{T})$. The application of this formula to $\mathcal{T}$ leads to:
\begin{equation}
 s=\partial_\epsilon\ln(\Gamma^2)-\left(\frac{1-\dot{\Sigma}}{\epsilon-\epsilon_1-\Sigma}+\frac{1-\dot{\Sigma}}{\epsilon-\epsilon_2-\Sigma}+\mbox{H.c.}\right) \label{Seebeck_Formula}
\end{equation}
The relation between the scattering matrix and the Hamiltonian is simple in the case of a two-level noninteracting system:
\begin{equation}
 \mathcal{S}=1-i\Gamma \frac{1}{\epsilon-H-\Sigma} \rightarrow \frac{1}{\epsilon-H-\Sigma}=\frac{1-\mathcal{S}}{i\Gamma} 
\end{equation}
Combination of the above expression with Eq. (\ref{Seebeck_Formula}) gives:
\begin{equation}
 s=\partial_\epsilon\ln(\Gamma^2)-(1-\dot{\Sigma}) {\rm Tr}\left(\frac{1-\mathcal{S}}{i\Gamma}\right)+(1-\dot{\Sigma}^\star) {\rm Tr}\left(\frac{1-\mathcal{S}^\dagger}{i\Gamma}\right)
\end{equation}
where the star symbol denotes the complex conjugate. Then we may rewrite $s$ as:
\begin{equation}
 s=\partial_\epsilon\ln(\Gamma^2)- 2\frac{\dot{\Gamma}}{\Gamma}+(1-\dot{\Sigma}) {\rm Tr}\left(\frac{\mathcal{S}}{i\Gamma}\right)-(1-\dot{\Sigma}^\star) {\rm Tr}\left(\frac{\mathcal{S}^\dagger}{i\Gamma}\right)
\end{equation}
which reduces to Eq.~(\ref{Seebeck2}). This expression has been obtained because the scattering matrix and the Hamiltonian have the same size (hence the interest in using the equivalent 2-level system); it corresponds to a system for which, in the continuum limit and at low Fermi energies, one deals with a universality class different from that of the bulk. We stress that Eq.~(\ref{Seebeck2}) constitutes an important result upon which all the subsequent statistical analysis is based.

\section{Statistical analysis} 

\subsection{Statistics of the Seebeck coefficient}
In this work, the scattering matrix ${\mathcal S}$ is a random variable, which we assume to be uniformly distributed; as such, ${\mathcal S}$ belongs to circular orthogonal ensembles (COE) \cite{Mehta}. The use of the circular ensemble is based on \textit{the equal a priori probability ansatz}\cite{Baranger2,Jalabert}. This is a natural choice when there is no reason to privilege any scattering matrix, in which case the mean of the distribution is $\langle\mathcal{S}\rangle =0 $. For a general case, obtained for different parameters than those leading to CE \cite{Abbout1,Abbout2}, we obtain a more general distribution called the Poisson kernel uniquely determined by its non vanishing mean scattering matrix $\langle \mathcal S \rangle \neq 0$ \cite{Mehta}. Since it is always possible to define a new unitary matrix uniformly distributed using matrix transformations \cite{BrouwerLorentz,Forrester}, the subsequent calculations are developed considering only the circular ensemble. 

The statistics of the Seebeck coefficient thus follows:

\begin{equation}
 \mathcal{P}_\epsilon(s)=\int \delta\left\{s-\frac{\alpha}{2}{\rm Tr}\left(\frac{e^{i\Theta}{\mathcal S}- e^{-i\Theta}{\mathcal S}^\dagger}{2i}\right)\right\} \delta_{\rm H} {\mathcal S}
\end{equation}

\noindent where  $\delta_{\rm H} {\mathcal S}$ denotes the Haar measure, which is invariant under the transformation ${\mathcal S}\rightarrow V{\mathcal S}V^T$ for any arbitrary unitary matrix $V$ (the superscript $^T$ denotes the transpose of a matrix). With $V=e^{i\Theta/2} {\mathds 1}_{2\times2}$, we obtain:
\begin{equation}
 \mathcal{P}_\epsilon(s)\!=\!\int \delta\left\{s-\frac{\alpha}{2}{\rm Tr}\left(\frac{\tilde{{\mathcal S}} - \tilde{{\mathcal S}}^\dagger}{2i}\right)\right\} \delta_{\rm H} \tilde{{\mathcal S}}=\frac{1}{\bar{\alpha}}  \mathcal{P}_{\epsilon=\epsilon_0}(s/\bar{\alpha}) \label{Invariance}
\end{equation}
\noindent where $\bar{\alpha}=\alpha/\alpha_0$ with $\alpha_0=\alpha(\epsilon_0)$. Equation~(\ref{Invariance}) shows that the general probability distribution of the Seebeck coefficient at arbitrary energy can be deduced \emph{directly} from the simpler one obtained at $\epsilon_0$. Note that for ease of notations, we retain ${\mathcal P}$ as a \emph{generic} notation for all the probability distributions in the subsequent parts of the article.

\subsection{Joint probability distribution function}
A similar analysis of the transmission $\mathcal{T}$ shows that its distribution does not depend on energy \cite{Note}. Since the transport coefficients $\mathcal {T}$ and $s$ are tightly related by the transport equations, the knowledge of the joint probability distribution function (j.p.d.f) $ \mathcal{P}_\epsilon(s,\mathcal{T})$ is required to study any observable depending on both $\mathcal{T}$ and $s$. The starting point for this is the rewriting of the scattering matrix using the following decomposition:
\begin{equation}\label{scatmatR}
 {\mathcal S}=R^T\left(
  \begin{array}{ c c }
     e^{i\theta_1} & 0 \\
     0 & e^{i\theta_2} 
  \end{array} \right) R
\end{equation}
where the rotation matrix $R$ is defined as:
\begin{equation}
 R=\frac{1}{\sqrt{2}}\left(
  \begin{array}{ c c }
     1 & -1 \\
     1 & \phantom{-}1 
  \end{array} \right) 
\end{equation}
\noindent The eigenphases $\theta_1$ and $\theta_2$ are independent and uniformly distributed in $[-\pi,+\pi]$ which is the consequence of ${\mathcal S}$ being taken from a COE with the left-right symmetry \cite{Mello}. 
Now, we analyze the j.p.d.f. of the Seebeck coefficient and transmission at the half-filling limit $\epsilon=\epsilon_0$ \cite{Note2}, with:
\begin{equation}
\mathcal{T}=\sin^2\left[(\theta_1-\theta_2)/2\right]
\end{equation}
\noindent and
\begin{equation}
s=\alpha_0\sin\left[(\theta_1+\theta_2)/2\right]\cos\left[(\theta_1-\theta_2)/2\right]
\end{equation}
The j.p.d.f. may now take the form:
\begin{equation}
\mathcal{P}_{\epsilon=\epsilon_0}(s,\mathcal{T})=\langle \delta\left(\mathcal{T}-\sin^2 u\right)\delta\left(s-\alpha_0\sin v\cos u\right)\rangle_{u,v}\label{Lengthy}
\end{equation}
\noindent where two independent and uniformly distributed variables $u=\frac{\theta_1-\theta_2}{2}$ and $v=\frac{\theta_1+\theta_2}{2}$ are introduced. As shown in the Supplemental Material, integration over these variables and use of Eq. (\ref{Invariance}) yield:
\begin{equation}
\mathcal{P}_{\epsilon}(s,\mathcal{T})=\frac{1}{\pi\sqrt{\mathcal{T}(1-\mathcal{T})}} \frac{1}{\pi\sqrt{\alpha^2(1-\mathcal{T})-s^2}} \label{JointProbability1}
\end{equation}
\noindent which constitutes one of the main results of this work since it shows the possible values of the couple $(S,\mathcal{T})$: $\mathcal{P}_{\epsilon}(S,\mathcal{T})$ is non-zero only if the following condition is satisfied:
\begin{equation}\label{cnstrnt0}
 \alpha^2 (1-\mathcal{T})-s^2>0, ~~~~~~ \mathcal{T}<1
\end{equation}
The relationship between the Seebeck coefficient and the transmission function is thus constrained by a parabolic law. This may be checked by sampling the ${\mathcal S}$  matrices belonging to the COE with the  left/right symmetry \cite{Explanation3}, from which the transport coefficients $\mathcal{T}$ and $s$ are numerically computed with Eqs.~(\ref{Seebeck}) and (\ref{Seebeck2}). The parabolic law is clearly shown on Fig.~\ref{Figure2} where the j.p.d.f. $\mathcal{P}_{\epsilon} (s,\mathcal{T})$ is plotted against $\mathcal{T}$ and $s$. 

Note that if we had used the Poisson Kernel distribution, which only changes the probability weight of each configuration of the system, the condition ~(\ref{cnstrnt0}) and the ensuing conclusions would still hold. 

\subsection{System performance}

From a thermodynamic viewpoint, thermoelectric systems connected to two thermal baths at temperatures $T_{\rm hot}$ and $T_{\rm cold}$, use their conduction electrons as a working fluid to directly convert a heat flux into electrical power and vice-versa with efficiency $\eta$. As for all heat engines, one seeks to increase either their so-called efficiency at maximum ouput power, $\eta_{P_{\rm max}}$, as discussed in numerous recent papers (see, e.g., Refs. \cite{Benenti,Saito,Sanchez,Balachandran} for mesoscopic systems and Ref. \cite{ApertetPRE1} for fundamental questions related to irreversibilities) or simply maximize the efficiency $\eta$. For simple models in thermoelectricity, an expression of the maximum of $\eta$, is related to the figure of merit of the system $ZT$ \cite{Ioffe,Goupil}: 
\begin{equation}\label{etamax}
\eta_{\rm max} = \eta_{\rm C}~\frac{\sqrt{1+ZT}-1}{\sqrt{1+ZT}+T_{\rm cold}/T_{\rm hot}}
\end{equation}

\noindent where $\eta_{\rm C}=1-T_{\rm cold}/T_{\rm hot}$ is the Carnot efficiency. 

Equation (\ref{etamax}) clearly shows that for given working conditions, $ZT$ is as good a device performance measure as the efficiency $\eta$ is. The maximum of the figure of merit $ZT$ is reached when the maximum of the power factor $\mathfrak{p}=\mathcal{T}s^2$ is reached. Satisfaction of this latter condition lies on the existence of the probability $\mathcal{P}_{\epsilon}(s,\mathcal{T}))$: Eq. (\ref{cnstrnt0}) implies that $\alpha\mathcal{T}(1-\mathcal{T})>\mathcal{T}s^2$, which in terms of power factor implies that $\mbox{max}(\alpha\mathcal{T}(1-\mathcal{T}))>\mathfrak{p}$, so that:

\begin{equation}\label{cnstrnt}
\alpha/4>\mathfrak{p}
\end{equation}

\noindent We thus find that $ZT$ reaches its maximum if the transmission $\mathcal{T}=1/2$.

\begin{figure}
 \begin{center}
   \includegraphics[scale=0.37]{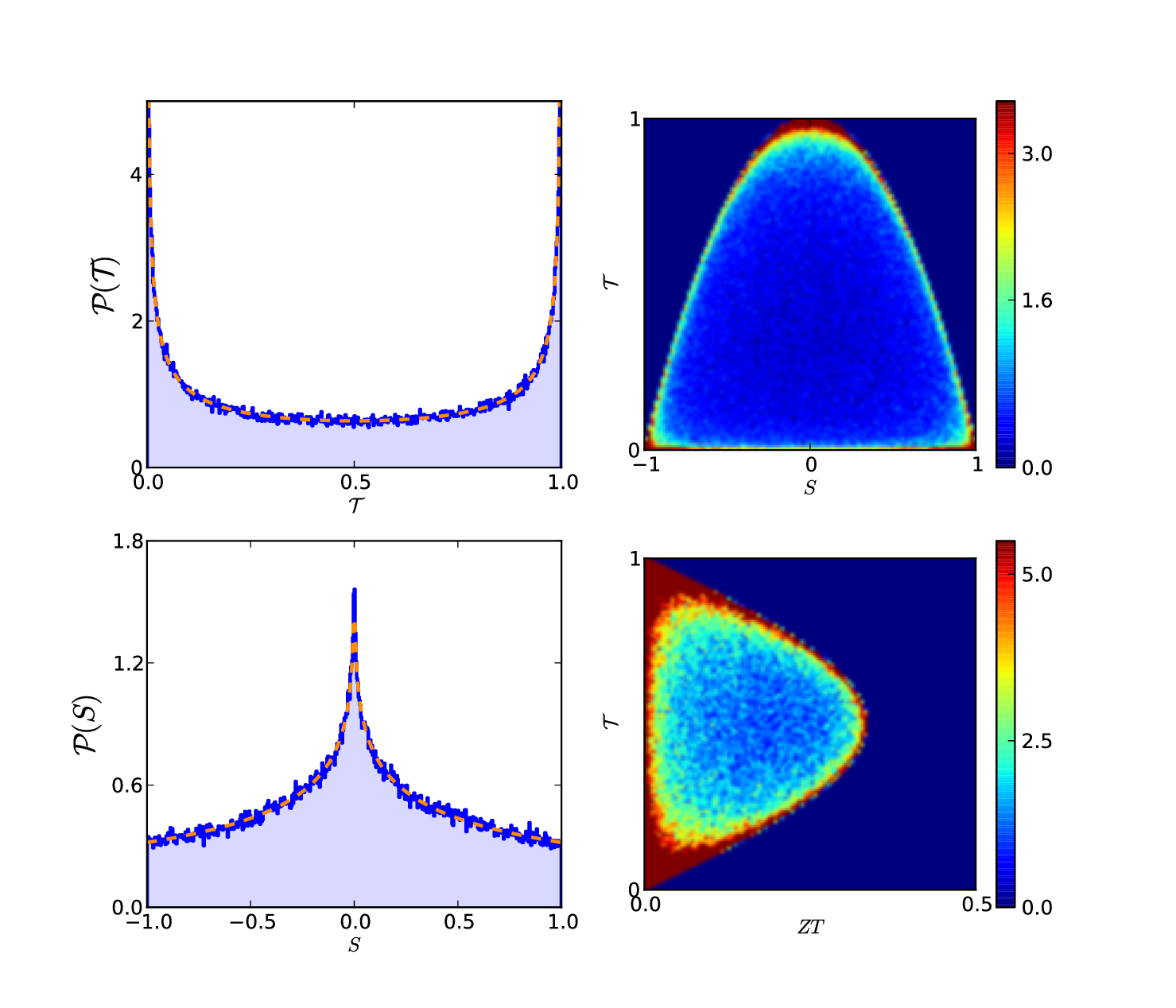}  
 \end{center}
\caption{(Color online) Probability densities:~1)Top left: $\mathcal{P}(\mathcal{T})$ for the transmission. 2) Bottom left: $\mathcal{P}(s)$ for the Seebeck coefficient. 3) Top right: j.p.d.f. of the couple $(s,\mathcal{T})$ at the half-filling limit.  4) Bottom right: j.p.d.f. of the couple $(\mathcal{T},ZT)$ at the half-filling limit. The blue curves in the left figures are obtained by sampling scattering matrices from circular orthogonal ensembles and the orange dashed curves represents the analytical result of the correponding physical observable. The matching is excellent.}
\label{Figure2}
\end{figure}

Physically, this means that the best peformance results from the conditions that make the system half-transparent; this is numerically checked on Fig.~(\ref{Figure2}). Note that the main assumption on the thermal properties of the system is that $\kappa$ is constant; this is justified on the condition that $\kappa_{\rm e} \ll \kappa_{\rm lat}$ \cite{Suekuni} ($\kappa_{\rm lat}$ is assumed to be constant). Relaxation of this assumption makes our system performance analysis applicable to the power factor only, but not to the figure of merit.

\subsection{Marginal distributions}
The marginal distributions $\mathcal{P}_{\epsilon}(\mathcal{T})$ and $\mathcal{P}_\epsilon(s)$ are obtained from the integration of $\mathcal{P}_{\epsilon}(s,\mathcal{T})$ over $s$ and $\mathcal{T}$ respectively:
\begin{eqnarray}
\mathcal{P}_{\epsilon}(\mathcal{T})&=&\frac{1}{\pi} \frac{1}{\sqrt{\mathcal{T}(1-\mathcal{T})}} \label{Transmission}\\
\mathcal{P}_{\epsilon}(s)&=&\frac{2}{\alpha\pi^2} \mathcal{K} \left(\sqrt{1-s^2/\alpha^2}\right)\label{Seebeckdist}
\end{eqnarray}
\noindent where $\mathcal{K}$ is the complete elliptic integral of the first kind. It is interesting to note that $\mathcal{P}_{\epsilon}(\mathcal{T})$ does not depend on the energy since it derives from the sole assumption that the scattering matrix is uniformly distributed (Dyson's CE) with the left/right spatial symmetry \cite{Mello}. As for all the physical observables with no energy derivative in their expression (shot noise for example), their distributions do not depend on the number of levels $N$ in the system: for all $N\ge$ 2, we obtain the same distribution, as it can be seen with the decimation method \cite{Abbout2,Abbout3} on the condition that ${\mathcal S} \in$ \{CE\}.

The case of the Seebeck coefficient is different: its distribution $\mathcal{P}_{\epsilon}(s)$ depends on the number of levels in the system since its expression contains an energy derivative. The only situation where the Seebeck coefficient, scaled with the appropriate parameter ($\alpha$ in our case), is independent on the number of levels is at the edge of the Hamiltonian spectrum \cite{Abbout1} where the result can be obtained by assuming the simpler two-level system considered as the minimal cavity for the two-mode scattering problem \cite{Abbout2}. The two-level model provides simple equations but yields general results, which apply to the $N$-level model at low density.

All the results presented so far in this article were obtained and discussed assuming a two-level system with left/right spatial symmetry and ${\mathcal S}\in$ \{CE\}. We must now see whether these hold when the assumption of left/right spatial symmetry is relaxed.

\subsection{Relaxation of the spatial symmetry assumption}
Assuming that there is no left/right symmetry, the two eigenphases $\theta_1$ and $\theta_2$ of Eq.~(\ref{scatmatR}) are no longer
independent, and we must consider a distribution of the form: $\mathcal{P}(\theta_1,\theta_2)\propto|e^{i\theta_1}-e^{i\theta_2}|$ \cite{Mehta}. The treatment is the same as that for the previous case, and while we obtain a different j.p.d.f. for the Seebeck and the transmission coefficients:
\begin{equation}
\mathcal{P}_{\epsilon}(s,\mathcal{T})=\frac{1}{\alpha\pi^2} \frac{1}{\sqrt{\mathcal{T}(1-\mathcal{T})}} \mathcal{K}\left(\sqrt{1-s^2/\alpha^2(1-\mathcal{T})}\right) \\ \label{JointProbability2}
\end{equation}
\noindent the mathematical constraint on the couples $(\mathcal{T},s)$ to obtain a non-vanishing joint probability distribution is exactly the same as Eq.~(\ref{cnstrnt}); this implies that the best system performance is obtained when it is half-transparent: $\mathcal{T}=\frac{1}{2}$.

The marginal distributions of $s$ and $\mathcal{T}$ are also different:
\begin{equation}
 \mathcal{P}_\epsilon(\mathcal{T})=\frac{1}{2\sqrt{\mathcal{T}}} \label{AssymetricTrans}
\end{equation}
\begin{equation}
\mathcal{P}_\epsilon(s)=-\frac{1}{\alpha \pi} \ln\left(\frac{|s|/\alpha}{1+\sqrt{1-s^2/\alpha^2}}\right)  \label{AssymetricSeeb}
\end{equation}
\noindent Equation (\ref{AssymetricTrans}) is consistent with results of Refs.~\cite{Jalabert,Baranger2} obtained with different methods and Eq. (\ref{AssymetricSeeb}) confirms the result of Ref.~\cite{Abbout1}. We see in both distributions, Eqs.~ (\ref{JointProbability1}) and (\ref{JointProbability2}), that the variables $(s,\mathcal{T})$ are not independent but it is interesting to note that if we define a new variable $X=s/\sqrt{(1-\mathcal{T})}$ then we have two independent variables, and the j.p.d.f. becomes multiplicatively separable:
\begin{equation}
 \mathcal{P}_\epsilon(X,\mathcal{T})=\mathcal{P}_\epsilon(\mathcal{T})\times \mathcal{P}_\epsilon(X)
\end{equation}
where we have $\mathcal{P}_\epsilon(X)=1/\pi\sqrt{\alpha^2-X^2}$ in the symmetric case and $\mathcal{P}_\epsilon(X)=\frac{2}{\alpha \pi^2} \mathcal{K}(\sqrt{1-X^2/\alpha^2})$ when the symmetry is relaxed; $\mathcal{P}_\epsilon(\mathcal{T})$,  given above, is the corresponding marginal distribution. It is worth  mentionning  that $S$ and $X$ also constitute a couple of independent variables.

\subsection{Statistics of the density of states}

The density of states in the two-level quantum dot may be expressed with the usual formula:
\begin{equation}
 \rho(\epsilon)=-\frac{1}{\pi} \Im {\rm Tr} G= -\frac{1}{\pi} \Im {\rm Tr}\frac{1}{\epsilon-H-\Sigma}
\end{equation}
which, using the scattering matrix, thus reads:
\begin{equation}
 \rho(\epsilon)=-\frac{1}{\pi} \Im {\rm Tr}\frac{1-\mathcal{S}}{i\Gamma}=-\frac{1}{\pi} \frac{1}{2i}{\rm Tr}\left(\frac{1-\mathcal{S}}{i\Gamma}- \frac{1-\mathcal{S}^\dagger}{-i\Gamma}\right)
\end{equation}
and simplifies to:
\begin{equation}
 \rho(\epsilon)=-\frac{1}{\pi}  \frac{1}{2i}\left[\frac{4}{i\Gamma}-{\rm Tr}\left(\frac{\mathcal{S}+\mathcal{S}^\dagger}{i\Gamma}\right)\right]
\end{equation}
so that
\begin{equation}
\frac{\delta \rho(\epsilon)}{\bar{\rho}(\epsilon)}=\frac{1}{2}  {\rm Tr}\left(\frac{\mathcal{S}+\mathcal{S}^\dagger}{2}\right)
\end{equation}
where $\bar{\rho}(\epsilon)=2/\pi\Gamma$. Here, we used the fact that for circular ensembles we do have $\bar{\mathcal{S}}=\bar{\mathcal{S}}^\dagger=0$ (where the over bar denotes the mean). 
The local density of states in the central system, when connected to the leads, reads \cite{Density}:
\begin{equation}
\rho(\epsilon)=-\frac{1}{\pi}\Im{\rm Tr}\frac{1}{\epsilon-\mathcal{H}_{\rm s}-\Sigma} \label{DensityConnected}
\end{equation}
\noindent which combined with Eq.~(\ref{SandHs}) yields:
\begin{equation}
 \frac{\delta \rho(\epsilon)}{\bar{\rho}(\epsilon)}=\frac{1}{2}{\rm Tr}\left(\frac{{\mathcal S}+{\mathcal S}^\dagger}{2}\right)
\end{equation}
where $\delta\rho=\rho-\bar{\rho}$. The analysis of this expression shows that the relative change in the density of states differs from the Seebeck coefficient, Eq.~(\ref{Seebeck2}), but the interesting result is that both $\delta \rho(\epsilon)/\bar{\rho}(\epsilon)$ and $s/\alpha$ have exactly the same distribution, which may be seen by using the invariance of the Haar measure under the transformation ${\mathcal S}\rightarrow-i {\mathcal S}$. We finally add to this set of variables the scaled Wigner time $\delta\tau_w/\alpha$ which is related to the time spent by a wavepacket in the scattering region: $\tau_w=-i\hbar{\rm Tr}( {\mathcal S}^\dagger \partial {\mathcal S}/\partial \epsilon)$ \cite{Abbout2}.

\begin{figure}
 \includegraphics[scale=0.4]{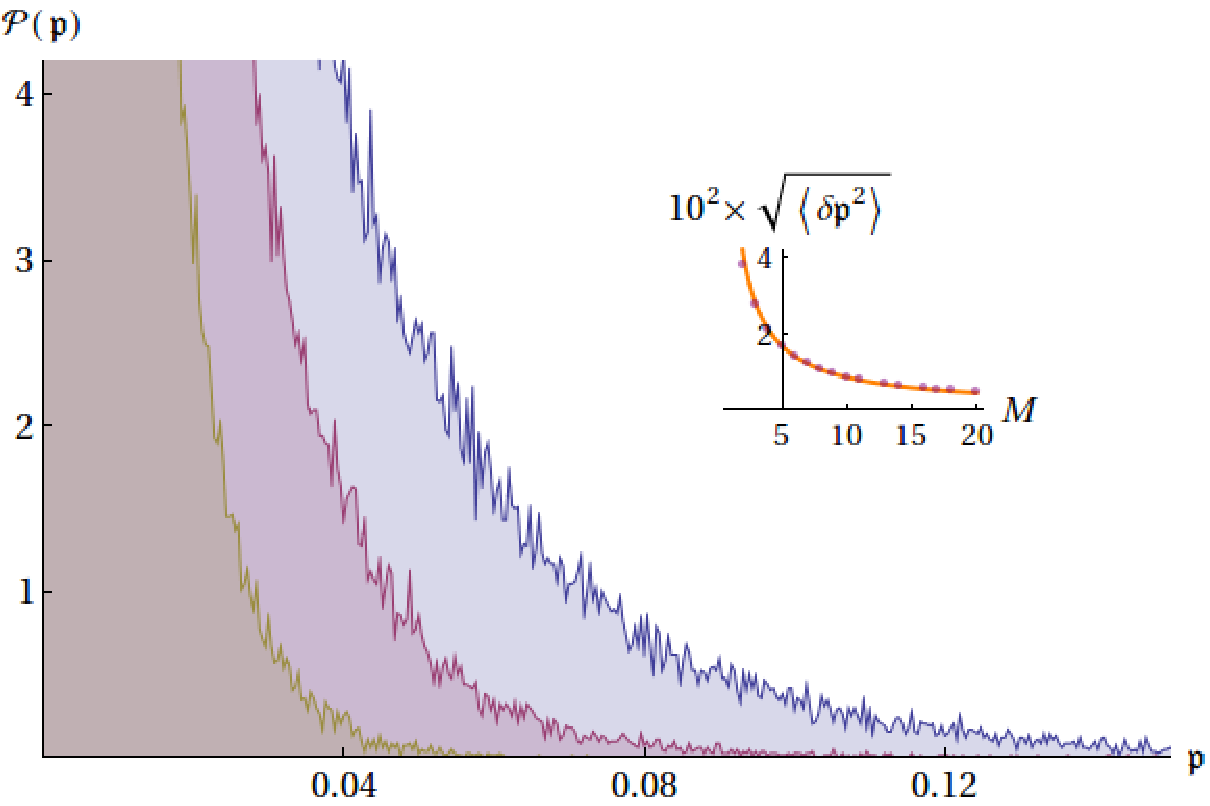}
\caption{(Color online) Probability density of the power factor $\mathfrak{p}$ for different values of the number of conduction modes: blue curve $M=4$, purple curve $M=8$, and brown curve $M=16$. The number of energy levels in the cavity equals the total number of modes. In the inset, the standard deviation of the power factor is shown as a function of $M$. The results are fitted with the function $\sqrt{\delta \mathfrak{p}^2}\sim 0.08/M$. The numerical results are obtained by sampling $2.1\times 10^5$ scattering matrices.}
\end{figure}
  
\section{Generalization to $M$ modes of conduction}
Here, we generalize and investigate the performance of a system with a large number of conduction modes $M$. We concentrate on systems with $2N$ levels connected to $2M$ independent and equivalent leads ($M$ on the left side and $M$ on the right side). To facilitate both the numerical and computational works, we make use of the model of minimal chaotic cavities and we concentrate on the edge of the Hamiltonian spectrum~\cite{Abbout2}, so that we can set $N=M$, and obtain results which are the same as those which would be produced for the general case $N>M$ (at the spectrum edge). We also assume no spatial symmetry.
\begin{figure}
\includegraphics[scale=0.4]{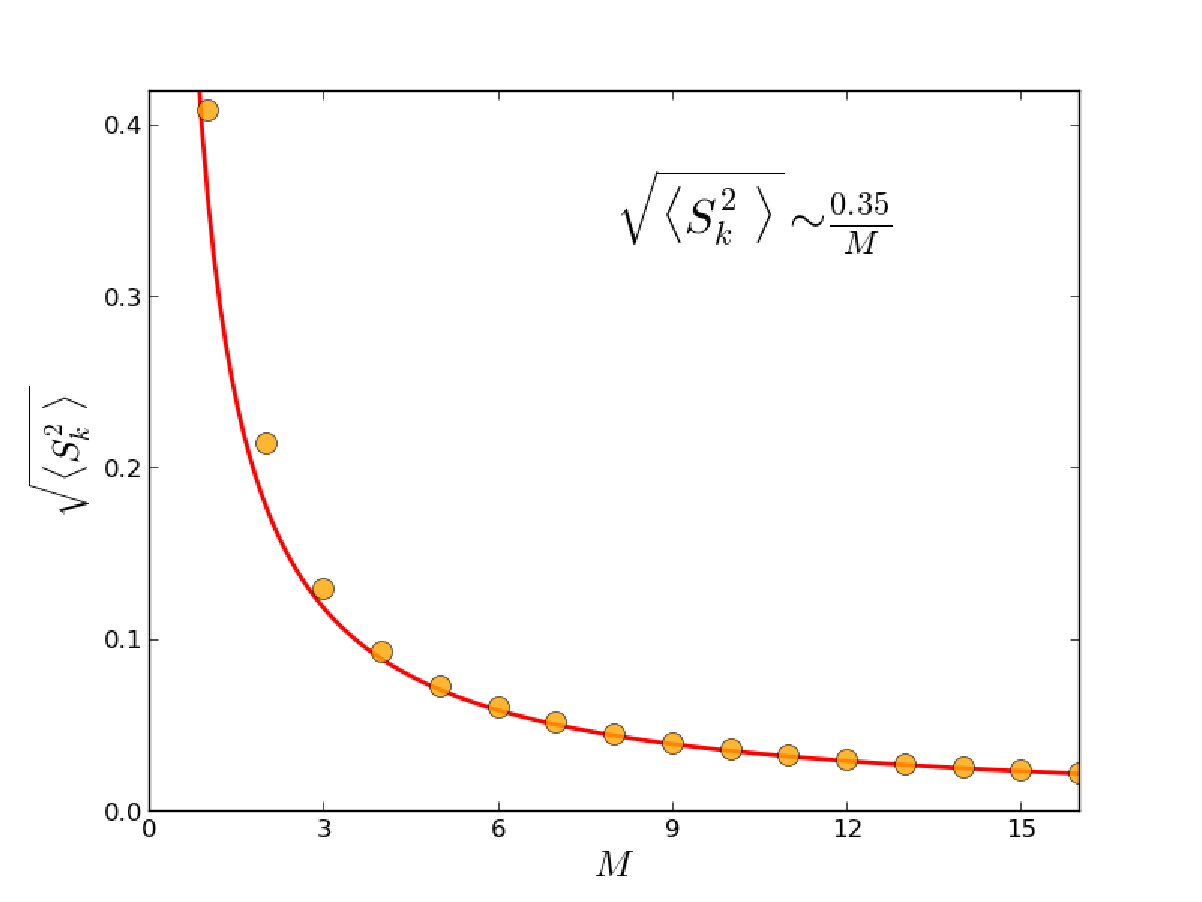} 
\caption{(Color online) Standard deviation of the Seebeck coefficient as a function of the number of modes for a minimal chaotic cavity ($M=N$). The Seebeck coefficient scales as $1/M$ for large numbers of modes.}
\label{FigureOfStd}
\end{figure}

The form of the scattering matrix which is now of size $2M\times2M$ remains unchanged [see Eq.~(\ref{SandHs})], and the transmission is given by $\mathcal{T}={\rm Tr}(t t^\dagger)$. For large values of $M$, the shape of the p.d.f. of the transmission tends to that of a Gaussian distribution for the typical values of $\mathcal{T}$ around its mean, which for $M$ large is given by $\bar{\mathcal{T}}\sim\frac{M}{2}$. The variance of this Gaussian is equal to $\langle\delta\mathcal{T}^2\rangle$=1/8 which does not depend on $M$ because of the universal character of the conductance fluctuations \cite{Beenakker}. It is worth mentionning that the tail of the distribution, which describes the atypical values of the conductance has a non-Gaussian form \cite{Pierpaolo}. At first glance, this scaling of the mean transmission seems to favour an enhancement of the power factor and hence of the figure of merit; however, our study of the Seebeck coefficient yields a different conclusion.

The Seebeck coefficient may be expressed by writing the derivative of the scattering matrix and using the expression $s=2 \Re({\rm Tr} (\dot{t}t^\dagger)/ \mathcal{T})$ (where the dot refers to the energy derivative and $\Re$ is the real part); then it is computed numerically assuming modes with a self-energy $\Sigma=\epsilon/2-i\sqrt{1-(\epsilon/2)^2}$ \cite{Sasada}. Still considering the half-filling limit since the generalization to arbitrary energy presents no particular difficulty, we find that the mean value of the Seebeck coefficient is always zero: $\langle s\rangle=0$ because the probability to obtain a positive thermopower is the same as that for a negative one. The most interesting result is the standard deviation which appears to scale as: $\langle\delta s^2\rangle^{1/2}=O(1/M)$. This clearly demonstrates that increasing the number of conduction modes in the system yields a decrease of the thermopower as shown in Fig.~\ref{FigureOfStd} where the standard deviation of thermopower is represented as a function of the number of modes on each side (left and right). The numerical results were obtained by sampling scattering matrices of size $2M\times2M$ from the circular orthogonal ensemble and computing a histogram from which the standard deviation was obtained.

\begin{figure}
\includegraphics[scale=0.35]{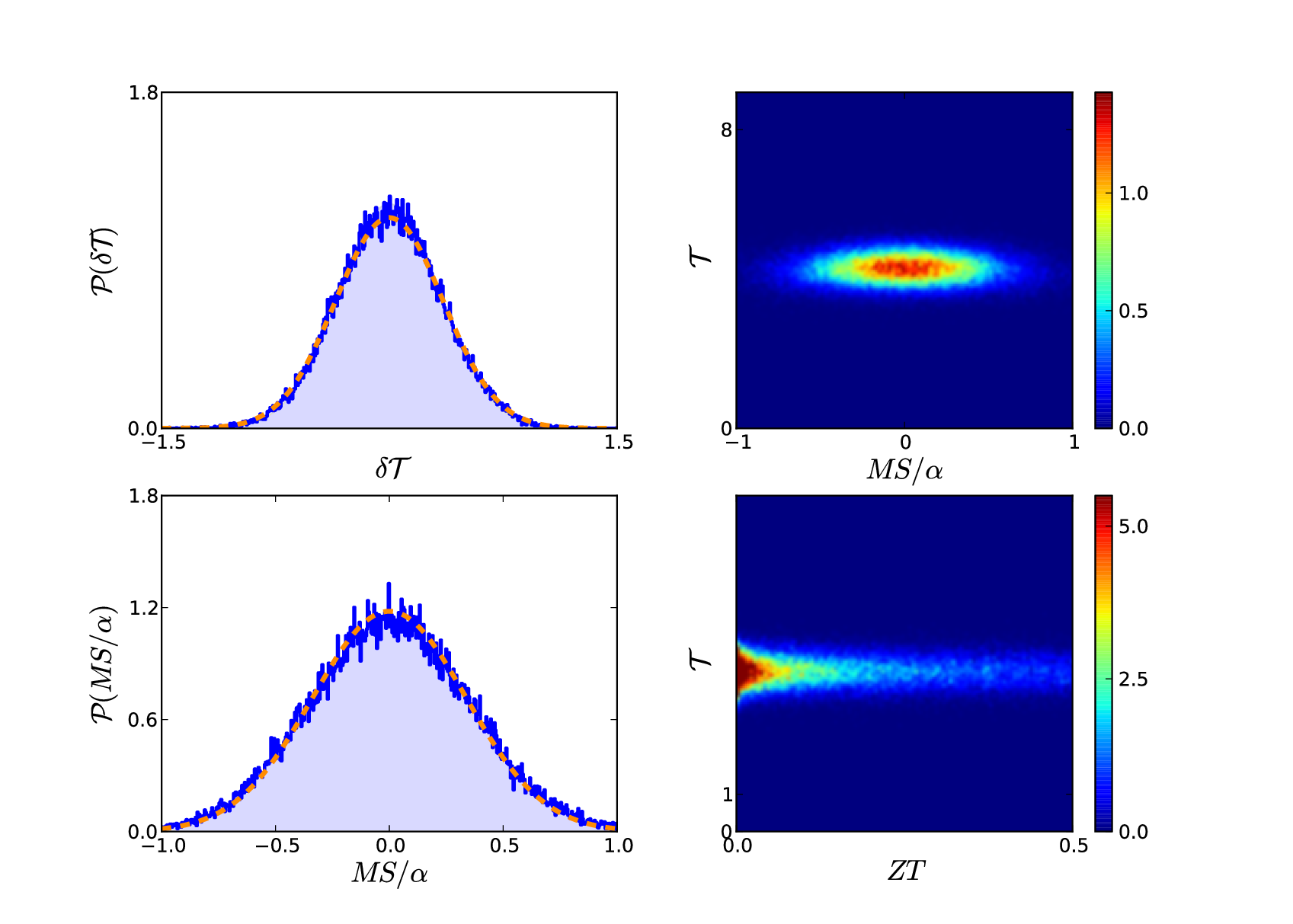}
\caption{Orange dashed line: Gaussian fit of the distributions $\mathcal{P}(\mathcal{T}-\bar{\mathcal{T}})$ compared to the numerical result (blue curve). The agreement is good.  $\mathcal{P}(M S/\alpha)$ (left panels), and j.p.d.f. for the couples $(\mathcal{T},ZT)$ and $(\mathcal{T}, MS/\alpha)$.}
\label{GaussFit}
\end{figure}

Since $\mathfrak{p}\propto s^2$, the lowering of the system performance induced by a decrease of the thermopower, which is faster than the increase due to conductance, is a reflection of the fact that the typical power factor is thus estimated to scale as $1/M$. We deduce from this analysis that the distributions $\mathcal{P}(\mathcal{T}-\bar{\mathcal{T}})$ and $\mathcal{P}(M s/\alpha)$ do not depend on $M$ for large values of $M$. In Fig.~\ref{GaussFit}, we see a very good agreement between the distribution $\mathcal{P}(M s/\alpha)$ and the Gaussian centered on zero. We also see from the j.p.d.f of $ZT$ and $\mathcal{T}$ that high values of the figure of merit are obtained for $\mathcal{T}\sim M/2$, which generalizes the result obtained for the two-level system.          

\section{Lattice model}
\begin{figure}
\includegraphics[scale=0.4]{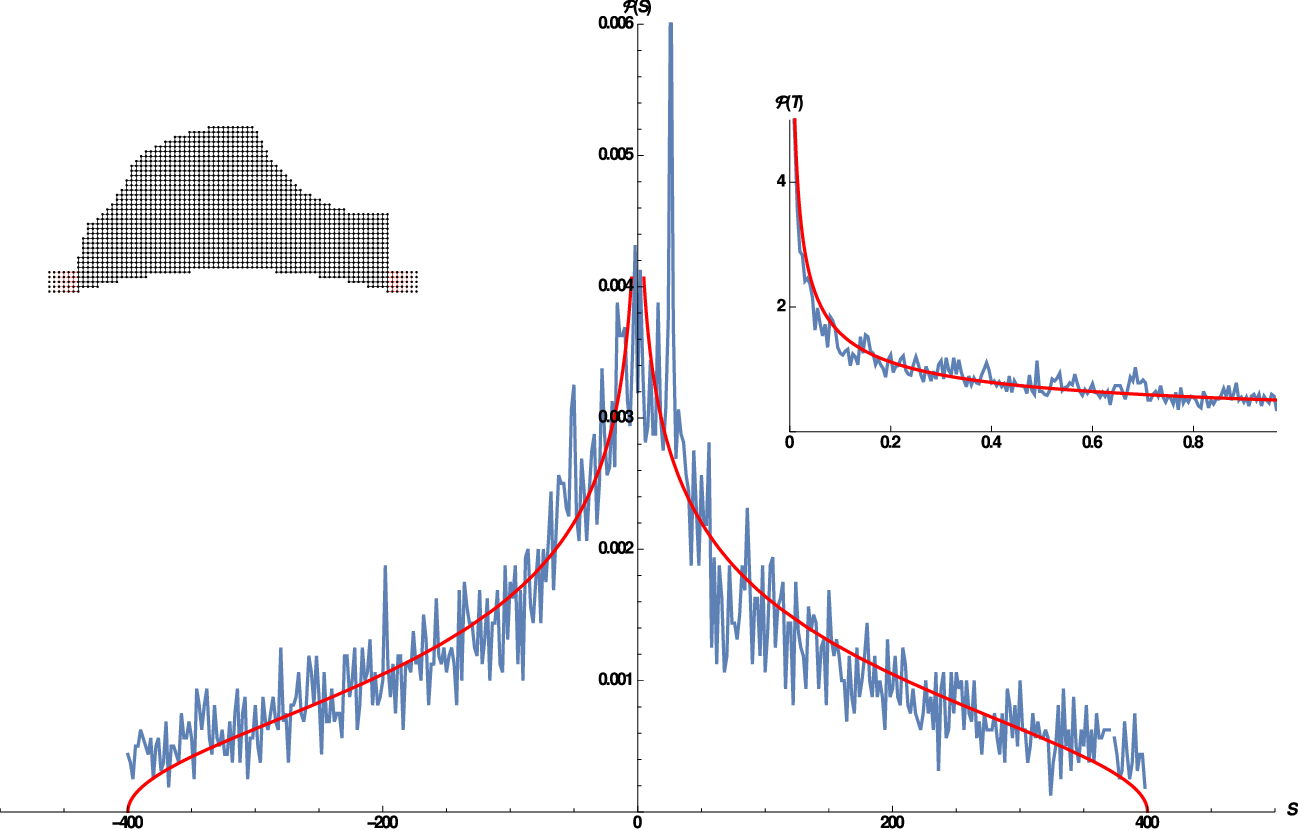}
\caption{Statistics of the Seebeck coefficient (Blue curve) in a realistic cavity simulated on a lattice. The sample is truncated and the very large values of $S$ were ommited due to numerical derviation over energy. (Red line), represents simulation using Eq.~(\ref{AssymetricSeeb}) with $\alpha$ as an ajustement parameter. The kind of cavity we used  is shown in the left inset. The law $\mathcal{P}_\epsilon(\mathcal{T})=\frac{1}{2\sqrt{T}}$ is verified in the right inset. The width of the leads are $5a$ and the Fermi energy is $E=-3.43t$    }
\label{LatticeModel}
\end{figure}
The Seebeck distributions obtained using the $2\times 2$ Hamiltonian reproduce those of the $N\times N$ system's Hamiltonian with $N>2$ as a limit at the spectrum edge \cite{Abbout1,Abbout2}. Increasing the number of levels $N$ will push the distribution towards that of the bulk spectrum \cite{Abbout1} where the mean level spacing is decaying as $1/N$. So for a given $N$, the Fermi energy has to be lowered in order to draw away from the bulk spectrum towards the spectrum edge. On the other hand, the Fermi energy must remain significant; that is why systems with  a relatively small number of levels $N$ are very well suited to our approach. Indeed, The center of the Hamiltonian distribution is different from the Fermi energy and this difference is larger at low Fermi energies. This can be understood and verified easily in a Graph-Hamiltonian model since the Lorentzian distribtion can be numerically generated. 

The case of a lattice model is a more challenging situation. We wish to verify the distribution of the transport coefficients in a manner which can be fullfilled experimentally and not only numerically. For this purpose, we study the transport in a cavity connected to two semi-infinite uniform leads. The cavity is subject to a small potential, uniform inside the cavity, but its value is randomly chosen in a small interval. This potential may be generated by external gates. The Fermi energy is chosen small yet significant to allow a unique conducting mode in the leads. The simulation on such a lattice model can be done using the Kwant software\cite{Kwant}. For a suitable range of magnitude of the small random potential, the result of the statistics of the conductance verifies the law given in Eq.~(\ref{AssymetricTrans}) as it can be seen in the right inset of Fig.~\ref{LatticeModel}. Once the distribution of the conductance is verified, we look towards the distribution of the Seebeck coefficient. This coefficient is obtained using a numerical derivation of the transmission with respect to energy. This numerical procedure can be sensitive, especially where the transmission and its derivative vanish at the same time \cite{Abbout3}. That is why we truncate the sample of the Seebeck values and get rid of the very large results. We also vary slightly the shape of the cavity used in the simulation as done in Ref.~\cite{Baranger2} to avoid any spatial symmetries. The shape of the cavity we use is shown in the left inset of Fig.~(\ref{LatticeModel}). The result of the simulation is shown in Fig.~(\ref{LatticeModel}) and was fitted with the analytical result of Eq.~(\ref{AssymetricSeeb}) with $\alpha$ as an ajustable parameter. 

\section{Discussion and concluding remarks}
We investigated the quantum thermoelectric transport of nanosystems made of two electron reservoirs connected through a low-density-of-states two-level chaotic quantum dot, using a statistical approach. Assuming noninteracting electrons but accounting for the energy dependence of the self energy and retaining its real part \cite{RemarkN}, and using the scattering matrix formalism, analytical expressions based on an \emph{exact} treatment of the chaotic behavior in such devices were obtained. Equation (\ref{Seebeck2}) in particular is an exact and general mathematical result, which could be obtained because the scattering matrix and the Hamiltonian have the same size (hence the interest in the treatment of the equivalent 2-level system). Analysis of the results provided the conditions for optimum performance of the system. Optimum efficiency is obtained for half-transparent dots and it may be enhanced for systems with few conducting modes, for which the exact transport coefficients probability distributions are found to be highly non-Gaussian. 

To end this article, we comment on the fact that our calculations are based on sampling of the scattering matrix using the equal a priori probability ansatz\cite{Baranger2,Jalabert}, instead of considering the Hamiltonian as a random matrix. This point is of interest as, instead of dealing with scattering matrices, one could of course follow a Hamiltonian approach with which integrations are performed over the eigenvalues instead of the eigenphases. The correspondence between the two formalisms is ensured by Eq. (\ref{SandHs}) and the corresponding distribution for the Hamiltonian, compatible with the circular ensembles is found, for a fixed $N$,\cite{RemarkN} to be Lorentzian.\cite{Abbout1,BrouwerLorentz} The equal a priori probability ansatz is therefore reflected in the relationship between the (center; width) of the distribution and the (real; imaginary) parts of the self-energy of the leads which ensures $\langle S\rangle= 0$.\cite{Abbout2,Abbout1} More precisely, the width of the eigenvalues needs to be equal to the broadening $\Gamma$ of the leads.\cite{Abbout1} Using a distribution different from the uniform one would of course imply that different weights are allocated to the possible configurations of the system, but while unlikely configurations would have no influence whatsoever on the outcome, unlike those which are possible, only these latter with larger weights would contribute to the maximization of the power factor, and the above conclusions would then remain essentially the same.


\begin{thebibliography}{9}
\bibitem{Explanation} By definition, at the edge of the spectrum, the levels are typically rare.
\bibitem{Note} To use the invariance of the Haar measure in the presence of transmission, one may better use this expression: $\mathcal{T}={\rm Tr}(C^1SC^2S^\dagger)$ with $C^1_{ij}=\delta_{i1}\delta_{j1}$ and $C^2_{ij}=\delta_{i2}\delta_{j2}$.
\bibitem{Explanation3} One would usually start by sampling the Hamiltonian first and then obtain the scattering matrix, but since generating Lorentzian ensemble uses unitary matrices, one may skip this step and sample directly ${\mathcal S}$. This simplification is possible here because the expressions we obtain involve only ${\mathcal S}$ in the minimal models.
\bibitem{Density} If the system is not connected to the leads, then $\rho(\epsilon)=-\frac{1}{\pi}\lim\limits_{\eta \to 0}\Im{\rm Tr}\frac{1}{\epsilon-\mathcal{H}_{\rm s}+i\eta}=\sum_{i=1}^2\delta(\epsilon-\epsilon_i)$, which yields $\bar{\rho}(\epsilon)=\frac{2}{\pi}\frac{\Gamma/2}{(\Gamma/2)^2+\Lambda^2}$. For 1D uniform leads, $\bar{\rho}(\epsilon)=\frac{\Gamma}{\pi}$.
\bibitem{RemarkN} This point is important since it is a condition to reach the university class of the spectrum edge by varying the Fermi energy to the continuum limit.
\end{thebibliography}
\end{document}